\documentclass[aip,jcp,amsmath,amssymb,reprint,floatfix]{revtex4-1}

\usepackage{graphicx}\usepackage{dcolumn}\usepackage{bm}\usepackage{siunitx}
\usepackage[utf8]{inputenc}
\usepackage[T1]{fontenc}
\usepackage{mathptmx}
\usepackage{hyperref}
\usepackage{xcolor}
\usepackage{soul}
\usepackage{pdfpages}
\usepackage{pgffor}

\makeatletter
\AtBeginDocument{\let\LS@rot\@undefined}
\makeatother

\definecolor{highlight}{RGB}{0, 0, 0}
\newcommand{\new}[1]{\textcolor{highlight}{#1}} 

\begin{document}

\title[Conformational analysis of tannic acid]{Conformational analysis of tannic acid: environment effects in electronic and reactivity properties}
\homepage[The following article has been accepted by The Journal of Chemical Physics. After it is published, it will be found at ]{https://aip.scitation.org/journal/jcp}

\author{Romana Petry}\email{romana.petry@lnnano.cnpem.br}
\affiliation{Center for Natural and Human Sciences, Federal University of ABC (UFABC), Santo André, 09210-580, São Paulo, Brazil}
\affiliation{Brazilian Nanotechnology National Laboratory (LNNano), Brazilian Center for Research in Energy and Materials (CNPEM), Campinas 13083-100, São Paulo, Brazil}

\author{Bruno Focassio}
\affiliation{Center for Natural and Human Sciences, Federal University of ABC (UFABC), Santo André, 09210-580, São Paulo, Brazil}
\affiliation{Brazilian Nanotechnology National Laboratory (LNNano), Brazilian Center for Research in Energy and Materials (CNPEM), Campinas 13083-100, São Paulo, Brazil}

\author{Gabriel R. Schleder}
\affiliation{Center for Natural and Human Sciences, Federal University of ABC (UFABC), Santo André, 09210-580, São Paulo, Brazil}
\affiliation{Brazilian Nanotechnology National Laboratory (LNNano), Brazilian Center for Research in Energy and Materials (CNPEM), Campinas 13083-100, São Paulo, Brazil}

\author{Diego S. T. Martinez}
\affiliation{Brazilian Nanotechnology National Laboratory (LNNano), Brazilian Center for Research in Energy and Materials (CNPEM), Campinas 13083-100, São Paulo, Brazil}

\author{Adalberto Fazzio}\email{adalberto.fazzio@lnnano.cnpem.br}
\affiliation{Center for Natural and Human Sciences, Federal University of ABC (UFABC), Santo André, 09210-580, São Paulo, Brazil}
\affiliation{Brazilian Nanotechnology National Laboratory (LNNano), Brazilian Center for Research in Energy and Materials (CNPEM), Campinas 13083-100, São Paulo, Brazil}


\begin{abstract}
Polyphenols are natural molecules of crucial importance in many applications, of which tannic acid (TA) is one of the most abundant and established.
Most high-value applications require precise control of TA interactions with the system of interest. However, the molecular structure of TA is still not comprehended at the atomic level, of which all electronic and reactivity properties depend. 
Here, we combine an enhanced sampling global optimization method with density functional theory (DFT)-based calculations to explore the \new{conformational space} of TA assisted by unsupervised machine learning visualization, and then investigate its lowest energy conformers.
We study the external environment's effect on the TA structure and properties.
We find that vacuum favors compact structures by stabilizing peripheral atoms' weak interactions, while in water, the molecule adopts more open conformations. 
The frontier molecular orbitals \new{of the conformers with lowest harmonic vibrational free energy} have a HOMO-LUMO energy gap of 2.21 (3.27) eV, increasing to 2.82 (3.88) eV in water, at the DFT generalized gradient approximation (and hybrid) level of theory. 
Structural differences also change the distribution of potential reactive sites. 
We establish the fundamental importance of accurate structural consideration in determining TA and related polyphenols interactions in relevant technological applications.
\end{abstract}

\maketitle

\section{\label{sec:intro}Introduction}
Polyphenols are biomolecules that participate in crucial biological processes, mainly found as secondary metabolites of plants, abundant in structures such as leaves, grains, and fruits. They comprehend a wide range of compounds structurally composed of more than one phenolic motifs, either in an oligomeric form or displayed in multiple monomeric forms \cite{platPlyphenol}. There is an immense industrial and technological interest in these molecules due to their attractive properties, including antioxidant, anti-tumor, antibacterial, and anti-inflammatory, besides their advantageous physicochemical properties \cite{Natural_poly_drug}. Polyphenols have great potential to interact with other molecules or surfaces via different mechanisms. It can also be structurally modified through straightforward chemical reactions \cite{poly_interfaces}, which made them suitable molecular platforms to a large range of applications, such as in environmental, material, and medical sciences \cite{biosorbenttannin,poly_materials,paclitaxel}.

\begin{figure*}[!ht]
\centering
\includegraphics[width=\linewidth]{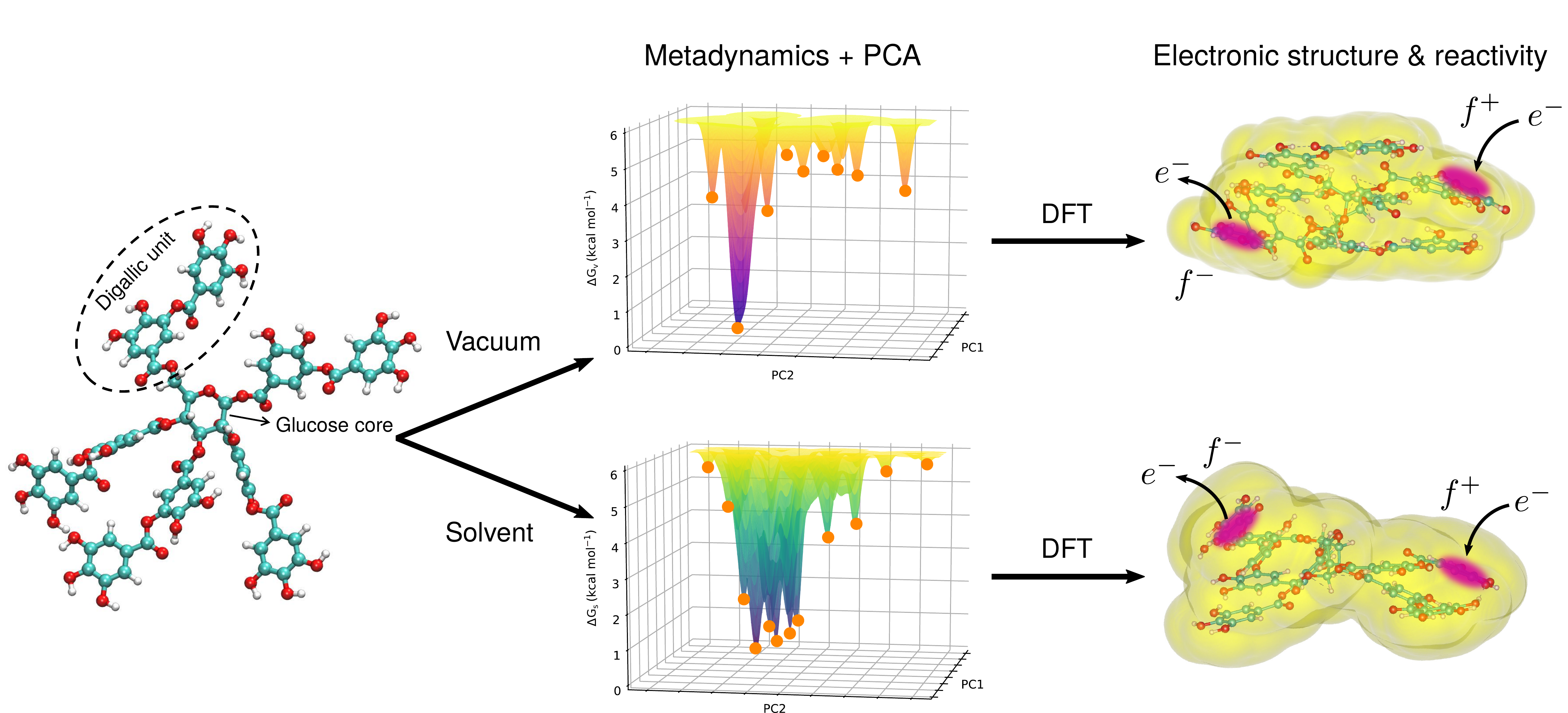}
\caption{Schematic workflow representation. Metadynamics is performed to find minimum free energy structures of tannic acid both in vacuum and in water solvent environments. The free energy surface is visualized by unsupervised machine learning methods. Different conformers have electronic and reactivity properties evaluated by DFT.}
\label{fig:tannic_initial_structure}
\end{figure*}

Tannic acid (TA) is a highly abundant and commercially available polyphenol with well-established applications in industrial sectors as diverse as food additives, flame-retardant, and surface coating \cite{fireretardant,food_tannins,PET_coating}. Its molecular structure consists of a decagalloyl glucose ($\rm{C}_{76}\rm{H}_{52}\rm{O}_{46}$), which is composed of five digallic acid units ester-linked to a glucose core as shown in  Fig. \ref{fig:tannic_initial_structure}. Besides the valuable biological activities, TA also presents favorable physicochemical properties such as high solubility, binding capacity to different molecules, metal ions coordination, and a great reducing and radical scavenging ability. This made it widely applied in the development of novel technologies.\cite{shin2018targeting,meetal_tannic_nanoteranostic,Radical}.     
In this context, TA has been used to integrate drug-delivery systems by binding to drugs' molecules or even functionalizing nano-formulated drug systems, improving its solubility, targeting, circulation time, therefore, its therapeutic efficiency \cite{shin2018targeting,Shen_TAFe}. TA also plays an important role in materials development. It is used as a building block, functionalization, or cross-linking compound of polymeric materials (hydrogels, membranes, films, micelles, nanoparticles) \cite{SHIN2019170,Hidrogel}. In the nanotechnology field, different TA properties are explored in nanomaterials synthesis, for example, for 2D materials exfoliation  \cite{PENG2020123288}, reduction of metallic ions \cite{ACSFe_TA}, and for reduction of graphene oxide \cite{GO_reduction,Graphene_TA}.

Given the significant properties of polyphenols, the fundamental understanding of their structural and electronic properties has a crucial role in enabling and ultimately controlling their applications. The knowledge of the molecule's structure determines its reactivity and electronic properties, consequently assessing its suitability for different technologies. Therefore, its determination is the first essential step.
Despite TA's industrial and technological relevance, few studies employ computational methodologies to understand its properties and interactions in different applied systems \cite{Chitosan, tannic_DSMO,lead,lukevs2015solvent}. The detailed conformational analysis and structural optimization of large and flexible molecules, such as TA, consist of a computationally costly task since they span a large conformation space with several local minima. However, to overcome this challenge, detailed configurational and electronic analysis of this molecule is needed, beyond usually employed structural approximations.

In this work, we combine metadynamics analysis with \textit{ab initio} calculations to explore TA's conformational space and compute its electronic properties and reactivity in different environments. Based on TA's biological relevance and applications, we obtained its global minimum structures in gas-phase and water, also considering a range of local minima conformers inside a $6\;\rm{kcal}\,\rm{mol}^{-1}$ energy window above those minima. The comparison between vacuum and solvent environments provide insightful information regarding the solvent effects on the molecules' conformational space and the resulting properties. The environment significantly influences TA's geometry, mainly over the weaker interactions between digallic acid units. The effects in the structure are also reflected in the electronic properties. We find that the lowest-energy configuration in solvent-ambient presents a highest occupied--lowest unoccupied molecular orbitals (HOMO-LUMO) energy gap 0.61 eV larger than the vacuum-ambient structure. The structural and electronic changes influenced in the molecule's reactive sites distribution, which reinforce the importance of accurate structural analysis using computational methodologies for prospecting and tuning TA applications.

\section{\label{sec:comput_methods}Computational Methods}

\subsection{Conformational sampling}
The exploration of the free energy surface (FES) of a molecular system is essential to obtain optimized and stable structures. This process involves sampling numerous conformations and calculating the total and free energies of each configuration.

We employed the GFN2-xTB \cite{GFN2} density-functional-based tight binding (DFTB) method combined with a Root-Mean-Square-Deviation (RMSD)-based metadynamics \cite{Grimme_2019} to sample TA's potential energy surface. 
The fundamental concept of the metadynamics method is applying a history-dependent potential that fills the minima of the potential energy surface (PES) over time, continually driving the structures away from previously sampled geometries and enabling to overcome energetic barriers\cite{MtaDParrinello,natureMetadynamics,PIETRUCCI201732}. Additionally, the system's representation is done by collective variables, consisting of the previous structures on the PES, represented by atomic RMSD in Cartesian space, Eq. \eqref{eq:rmsd}:
\begin{equation}
    {\rm RMSD}_i = \sqrt{\frac{1}{N} \sum_{n=1}^{N} (r_n^{i} - r_n^{\rm ref})^2 }\;, \label{eq:rmsd}
\end{equation}
\noindent where $N$ is the number of atoms, $r_n$ is the position of the $n$th atom of the $i$th conformer and $r_n^{\rm ref}$ is the position of the $n$th atom of the reference structure.
The method is implemented in the Conformer-Rotamer Ensemble Sampling Tool (CREST) as available in the xTB code, capable of exploring compound's conformers in large PES \cite{crest,XTB_Method}.

The conformational analysis of TA started from a molecular structure without optimization, Fig. \ref{fig:tannic_initial_structure}, obtained from the Pubchem database \cite{Pubchem}. The global optimization was performed at 298.5K in both conditions, gas-phase and considering a water solvent media by employing the Generalized Born model with the solvent-accessible Surface Area (GBSA) model.\cite{GBSA} The conformer search provided lowest-energy structures in each condition, and also its conformers with energies up to $6\;\rm{kcal}\,\rm{mol}^{-1}$ ($0.26\;\rm eV$) above the minimum.
The thresholds for conformers' identification, which comprehend atomic RMSD, rotational constant, and $\Delta E$, were respectively 0.125 Å , 1\% (internally adjusted to 2\% in posterior steps), and $0.1\;\rm{kcal}\,\rm{mol}^{-1}$ ($4\;\rm meV$). Finally, to compute the conformers' free energy, we employed the coupled rigid-rotor-harmonic-oscillator (RRHO) approach as implemented in the xTB code \cite{GFN2,Spicher2021}. \new{Therefore, the free energy and the FES in this work include harmonic vibrational free energy contributions to the potential energy.}

For visualization and exploration of the numerous conformers provided by the CREST metadynamics, we created a two-dimensional map of the conformer space based on the RMSD information \cite{Isayev2015,De2016}. The different molecules' conformations were numerically represented (featurized)\cite{Schleder_2019,Schleder2020,Giustino2021} through a many-body tensor representation (MBTR) \cite{huo2018unified_mbtr,Himanen2020} including one, two, and three-body terms, i.e., taking into consideration the atomic number, inverse of the distance, and cosine of the angle, respectively. In this way, each conformer was represented by 2700 features. We then performed principal component analysis (PCA) to reduce the dimensionality of this large feature space to two-dimensions, using the conformer space of both media simultaneously. Each conformer is represented by a point in this 2D space and color-coded according to the RMSD, calculated using Eq. \eqref{eq:rmsd}.
The reference geometry was taken as the \new{free energy minimum} for the corresponding medium (vacuum and solvent). The conformers were aligned according to its reference, eliminating the contribution of rigid translations and rotations to the RMSD \cite{Melander2015}.

\subsection{Electronic structure}

TA structures with lowest free energy and potential energy as obtained by CREST were further optimized and had the electronic properties evaluated and compared via density functional theory (DFT) \cite{dft1964,dft1965} calculations, using the CP2K \cite{cp2kquickstep} code.
\new{Other representative structures of TA were also selected to evaluate the electronic properties by performing a hierarchical clustering of each ensemble structure obtained at the xTB level of theory, according to four features: the harmonic vibrational free energy, the structural RMSD, and the two first two principal components of the PCA on the MBTR descriptor. The selected conformers vary in relation to group, free energy and RMSD as presented in Fig. S2 in the Supplementary Material.} 

For DFT calculations, we used the revised Perdew, Burke and Ernzerhof (revPBE) generalized gradient approximation for the exchange-correlation term \cite{PBE,PhysRevLett.80.890}. A molecular optimized double-$\zeta$ Gaussian basis set plus polarization (DZVP-MOLOPT) was used, accompanied by Goedecker-Teter-Hutter pseudopotentials \cite{krack2005pseudopotentials}. The plane wave expansion energy cutoff was $60\;\rm Ry$ and the real-space grid cutoff was $500\;\rm Ry$. Furthermore, we employed the Grimme DFT-D3(BJ) method \cite{BJBJ, BJ_grimee,DFTD3} to account for dispersion interactions, shown to accurately describe molecular energies and geometries \cite{Schleder2017}. To account for solvation effects, the implicit self-consistent continuum solvation model (SCCS) \cite{Andreussi_SCCS} was applied for conformers previously obtained in this condition. The atomic positions were optimized until residual forces were smaller than $10^{-2}\;\rm{eV}\,\si{\angstrom}^{-1}$. The projected density of states (PDOS) and the relative energy alignment of HOMO, HOMO($-1$), LUMO, and LUMO($+1$) orbitals were analyzed in each condition (vacuum and solvent). The systems' eigenvalues were aligned according to a reference system, which we considered the optimized conformation of TA in vacuum. Finally, to evaluate reactivity changes, Fukui functions were calculated\cite{Fukui1952,Fukui1982,Parr1984,Ayers2000,ALLISON2013334}, analyzing differences in electron density when an electron is removed, Eq. \eqref{eq:fukuimenos}, or added, Eq. \eqref{eq:fukuimais}, to the molecule:
\begin{equation}
    {f^{-}} = \rho (\rm N_{e}) - \rho (\rm N_{e} - 1)\; \label{eq:fukuimenos}
\end{equation}
\begin{equation}
    {f^{+}} = \rho (\rm N_{e} + 1) - \rho (\rm N_{e})\;, \label{eq:fukuimais}
\end{equation}
where the electron density $\rho(N_{e})$, $\rho(N_{e} - 1)$, and $\rho(N_{e}+1)$  corresponds to the system with $N_{e}$, $N_{e}-1$, and $N_{e}+1$ electrons, respectively. Afterward, the condensed Fukui functions\cite{Roy1999,Roy2000} were calculated to analyze atomic reactivities:
\begin{equation}
    {f^{-}_n} = q_n({\rm N_e}) - q_n(\rm N_e - 1)\; \label{eq:cfukuimenos}
\end{equation}
\begin{equation}
    {f^{+}_n} = q_n({\rm N_e} + 1) - q_n(\rm N_e)\;, \label{eq:cfukuimais}
\end{equation}
where $q_n$ is the atomic charge of the $n$th site, which was computed through Mulliken population analysis.

\section{Results and discussion}

\begin{figure*}[!ht]
\centering
\includegraphics[width=\linewidth]{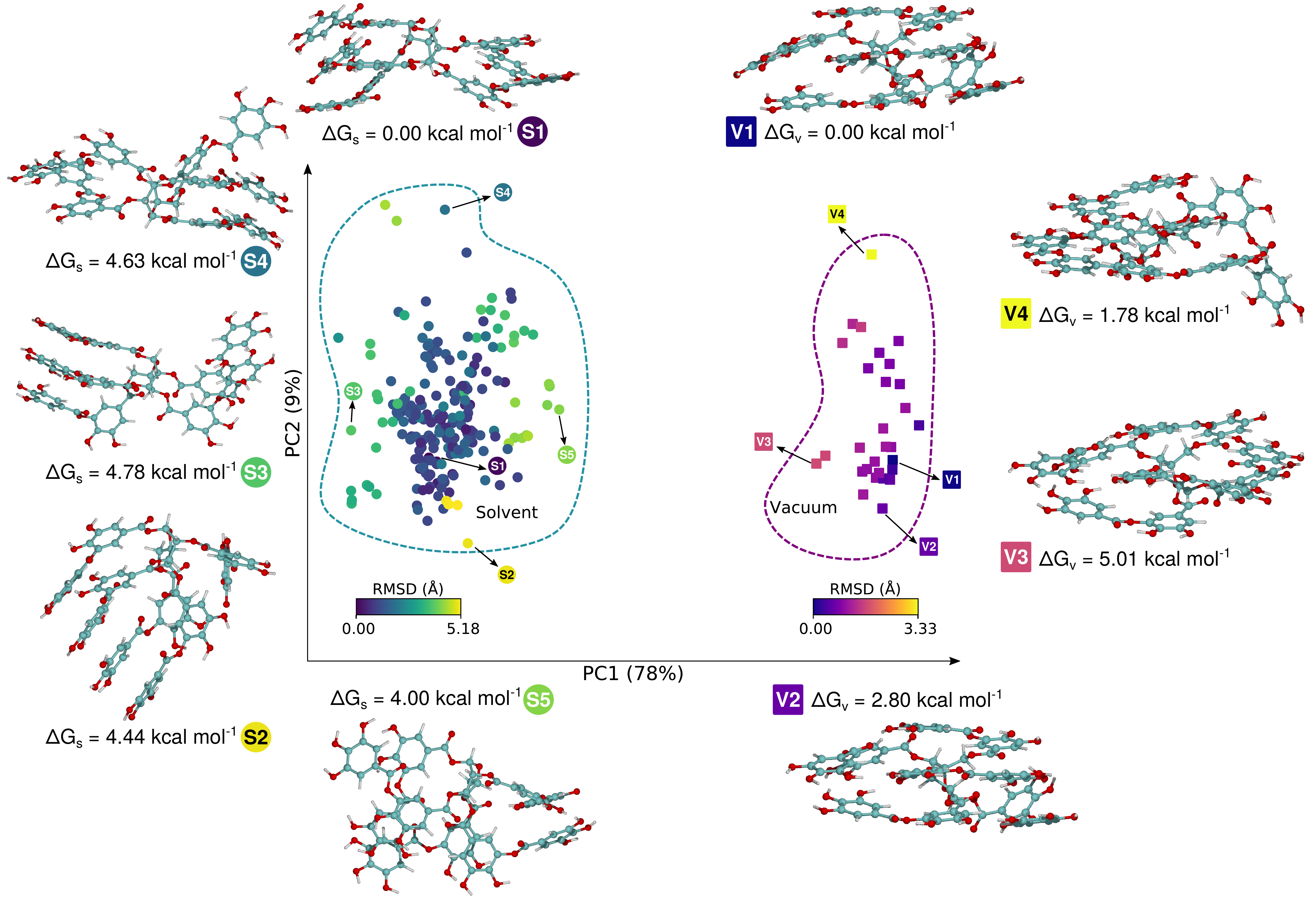}
\caption{Unsupervised principal component analysis (PCA) map of conformer configuration space. The TA's conformations were featurized with the MBTR representation and submitted to linear dimensionality reduction into a 2D feature space. Each square or circle represents a different geometry obtained from vacuum or solvent calculations, respectively. The color scales correspond to the RMSD of each structure relative to the global minimum of its respective calculation environment. Free energy minima structures (V1 and S1) and representative configurations are presented around the map. The configurations labeled as ``V'' or ``S'' were obtained in vacuum or implicit solvation, respectively. The free energy difference to the corresponding minima is given below the structures. The first and second principal components (PC1, PC2) corresponds to 78\% and 9\% of the total dataset variance.}
\label{fig:pca_analysis}
\end{figure*}

Detailed conformational analysis is essential to understand the structure-activity relationships of biomolecular systems. The different conformations can present significant changes in electronic properties, exposed sites for interaction with other systems, and even biological activities.

Due to the TA flexible structure, the FES representing its conformational space is expected to present various local energy minima. Such characteristics make structural optimization of this molecule challenging since different theory levels lead to distinct molecular structures. To avoid non-global minima structures, it is necessary to systematically explore the FES to search for structures with the lowest potential energies.
\textit{Ab initio} methodologies can efficiently evaluate local minimum areas of the FES with high accuracy but in its simpler forms are inefficient to sample larger areas of the FES \cite{Schleder_2019,MtaDParrinello}.
Optimized strategies are necessary to assess larger configuration space regions, overcoming energetic barriers between the local minima. 
We utilized a metadynamics method to find the global minimum configuration of TA in two different conditions, gas-phase (i.e., in vacuum) and implicit water solvation, evaluating the solvent's role in the final structure. We then evaluated its electronic and reactivity properties through \textit{ab initio} calculations. In Fig. \ref{fig:tannic_initial_structure} we present the workflow schematically.

We obtained a TA configuration ensemble for each condition, consisting of the structure with the lowest free energy and geometries $6\;\rm kcal\,mol^{-1}$ above this minimum.In vacuum, 29 different conformers were obtained for TA, and 212 conformers in solvent. 
To quantitatively evaluate all conformations' structural differences, we perform a data-driven combined principal component analysis that maximizes the configurations' variance into two axes.
Fig. \ref{fig:pca_analysis} show TA configurations with distinct RMSD projected in the principal components map. 
As expected, the conformations obtained in vacuum (squares) and in solvent (circles) exhibit significant structural differences and are completely separated in the two principal components. 
V1 and S1 represent the minima configurations of each condition, respectively. Among each group, the structural variations are numerically represented by the RMSD relative to the global minimum of the respective calculated condition, illustrated by the color scales.

\begin{figure*}[!ht]
\centering
\includegraphics[width=1.0\linewidth]{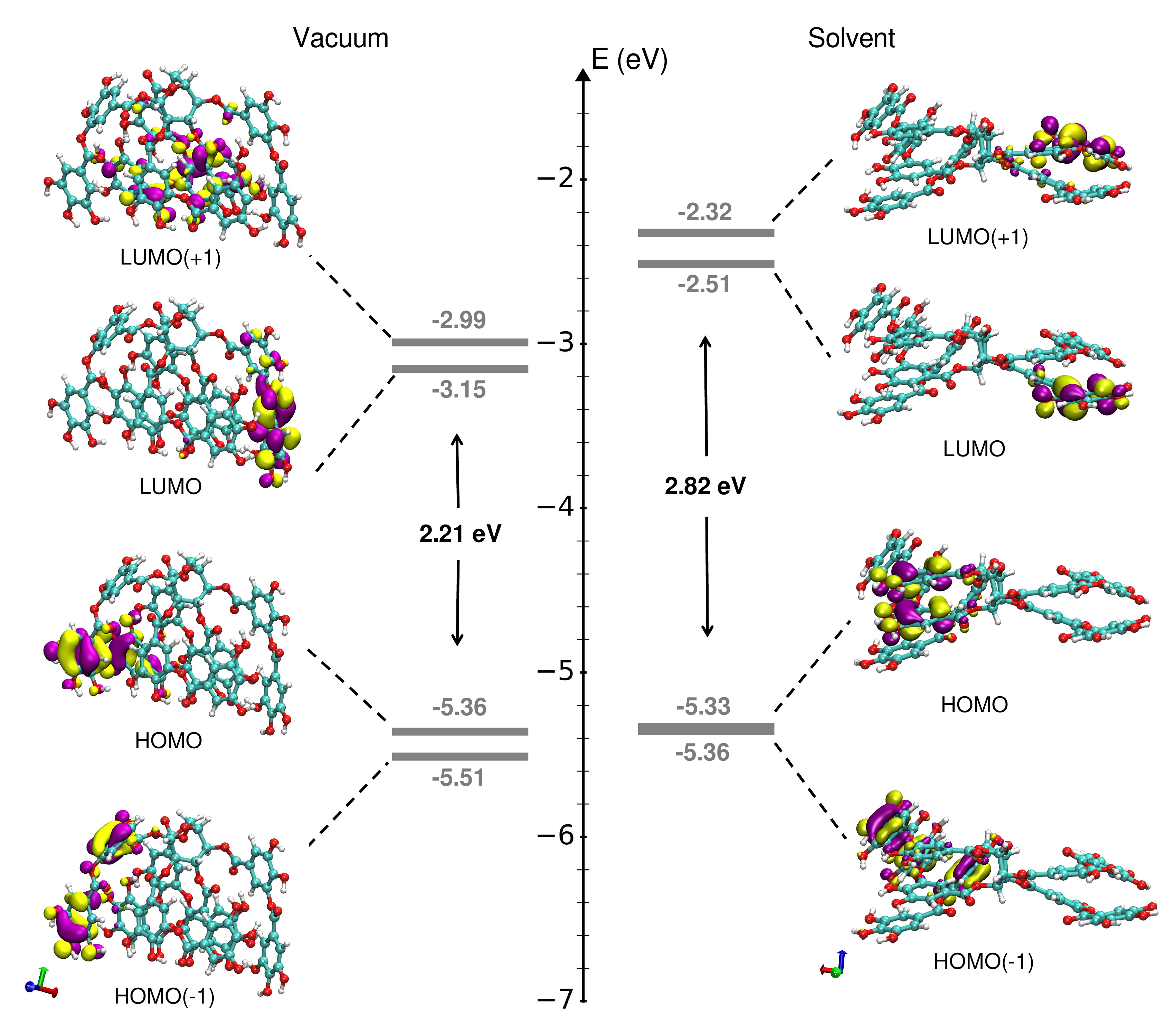}
\caption{Frontier molecular orbitals of free energy minima TA structures obtained in vacuum and implicit solvent: HOMO($-1$), HOMO, LUMO, LUMO($+1$), and the HOMO-LUMO difference. The eigenvalues were aligned according to the vacuum level obtained from the optimized vacuum TA calculation. Isosurfaces of $0.08\;\si{\angstrom}^{-3/2}$. Yellow and purple colors indicate opposite signs of the orbital wave function.}
\label{fig:wfc_homo_lumo}
\end{figure*}

The conformers differ mainly in the rotation of the digallic portions, their angulation in relation to the glucose core, and the number of intramolecular interactions. The solvated environment led to more open configurations than in vacuum, in which all digallic portions are closely interacting. 
In Fig. S8 of the Supplementary Material we show that the interaction energy of different intramolecular interactions, $\pi$-stacking and hydrogen bonds, between TA's digallic portions, decreases in the solvent environment.
The hydrophobic and electrostatic solvation forces comprehended in the GBSA solvent model compete with the intramolecular Coulomb and van der Waals forces \cite{Balancingforces}.
This competition effect is more pronounced for hydrogen bonds, whose interaction energy decreases by almost half in the solvent environment. Depending on the extension of these specific interactions, the solvent leads to different conformational equilibria, enabling that more open conformations present lower energies.

In the vacuum environment the intramolecular interactions do not have any competing effects and are more prevalent. In this condition, small structural changes presented more expressive energy differences in comparison with the minimum configuration. Most of the obtained configurations presented small values of RMSD, with energy differences ($\Delta{G}$) close to $6\;\rm kcal\,mol^{-1}$, see Fig. S1 in the Supplementary Material. Contrarily, in water, the solvation stabilization allows extensive structural changes, covering a broader RMSD space with a smoother energy distribution between conformations.

We now evaluate the impact of the structural differences in TA's electronic properties. The lowest free and potential energy structures obtained \new{from metadynamics calculations, and a range of different structures selected based on a hierarchical clustering approach (Fig. S2) to account for trends near the energetic minima \cite{Baldauf_2015,Ceriotti2019},} were used in further structural relaxation and electronic evaluation at the DFT level of theory. Table S1 \new{and Fig. S3} in the Supplementary Material summarizes these results.
Compared to DFT, most vacuum conformations at the GFN2-xTB level of theory are energetically close, indicating that the PES is flatter, a known feature of this level of theory \cite{crest}.

\new{In the following, we focus on the detailed electronic properties of the harmonic vibrational free energy minimum structures}. The results for other conformations can be found on the Supplementary Material, \new{Figs. S4 to S7}. Figure \ref{fig:wfc_homo_lumo} shows the eigenvalues and real-space wavefunction representation of the frontier molecular orbitals, indicating regions of possible electron donation (HOMO) or acceptance (LUMO). The projected density of states shows that the HOMO/LUMO orbitals are mainly formed by carbon and oxygen contributions, see Fig. S5. These orbitals are located at opposite sites of the molecule in both environments, mostly over the aromatic end rings. Interestingly, the electronic density distribution of the gallic acid molecule ---the smallest TA unit--- is mostly preserved in TA in both calculated conditions. This encourages the use of this molecule as a theoretical model for TA interaction studies. However, the contributions from other portions of TA can not be ignored. The higher number of atoms and intramolecular interactions make TA's HOMO-LUMO energy gap significantly smaller than gallic acid, see Fig. 3 and S9 in the Supplementary Material. Besides, differently than in gallic acid, the HOMO-LUMO gap of TA increases in the solvent due to the substantial structural rearrangement.

In the dielectric medium, TA presented a HOMO-LUMO energy gap of \SI{0.61}{\electronvolt} larger than in gas-phase. Different conformers also present this trend, available in the Supplementary Material, \new{Figs. S3 and S4}. To verify the trend between HOMO-LUMO energy gap in vacuum and solvent environments in a higher theory level, closer to experiments, we resorted to the hybrid functional of Heyd-Scuseria-Ernzerhof (HSE06) \cite{Heyd2003,Heyd2004,Heyd2006,Krukau2006} which resulted in an almost constant energy shift for both vacuum (HOMO-LUMO gap of 3.27 eV) and solvent (HOMO-LUMO gap of 3.81 eV). The difference between the vacuum and solvent HOMO-LUMO gap is 0.61 eV with HSE06, close to the previous value, confirming the observed trend and solvent effect. 

\new{The electronic calculations of the clusters' representative set of structures, available in Figs. S6 and S7, corroborate the influence of structural changes in electronic properties caused by the different environments. On average, vacuum conformations are more compact, having smaller HOMO-LUMO gaps with frontier orbitals located at opposite sites of the molecule, varying between portions of gallic acid and degree of delocalization. Conversely, solvent conformations are more open with larger electronic gaps when compared to vacuum. In each environment, few structures presenting smaller gaps can exist. Nonetheless, upon closer inspection, they show an individual lower-symmetry localized state inside the gap \cite{Fukui1982,woodward}, as indicated in Figs. S6 and S7.}

The results show that the environment significantly influences TA's geometry and, consequently, its electronic structure. The electronic density distribution is determinant of many properties, e.g., electronegativity, optical transitions, and redox potentials, and also of molecular systems' behavior. In this context, the frontier orbitals play a major role in molecules' interactions and reactivity. The alignment of these orbitals in relation to other systems will dictate the type and even pathways of reactions that they are involved in.\cite{FUJIMOTO1972177,woodward} The hardness, as introduced by \citet{Pearson}, is an example of a reactivity parameter directly related to frontier orbitals. It may be approximated to the difference between the HOMO and LUMO energies, and can be interpreted as the system's resistance to changes in the number of electrons. Therefore,  changes in the HOMO-LUMO gap observed between TA's structures in different environments are expected to impact molecules' reactivity.

The electronic Fukui functions are a concept used to study the local reactivity of molecules. They give information regarding how the loss or gain of electrons affects the spatial electronic density of the molecule \cite{Fukui1952,Fukui1982}. The condensed Fukui functions also provide insights about the relative nucleophilicity and electrophilicity, applied to atoms within the molecule\cite{Roy1999,Roy2000}. They are calculated from the electronic density difference when an electron is added or removed from the system, as given in Eq. \ref{eq:fukuimenos} to \ref{eq:cfukuimais}. TA $f^+$ and $f^-$ functions in vacuum and solvent are presented in Fig. \ref{fig:fukui} and their condensed forms in Tables S2 and S3 in the Supplementary Material. The Fukui functions predict the HOMO and LUMO sites as the most affected sites by electrophilic and nucleophilic reactions, respectively. The $f^+$ and $f^-$ are more delocalized in the TA configuration obtained in solvent than in vacuum, especially $f^-$ that appears in almost all digallic portions in Fig. \ref{fig:fukui}b, the condensed functions corroborate this result. This analysis indicate that the structural changes in this environment also affected the distribution of reactive sites in TA. Furthermore, $f^+$ densities are predominant in most of the molecule sites in both conditions, confirming the radical scavenging property of TA.    
\begin{figure}[!ht]
\includegraphics[width=1.0\columnwidth]{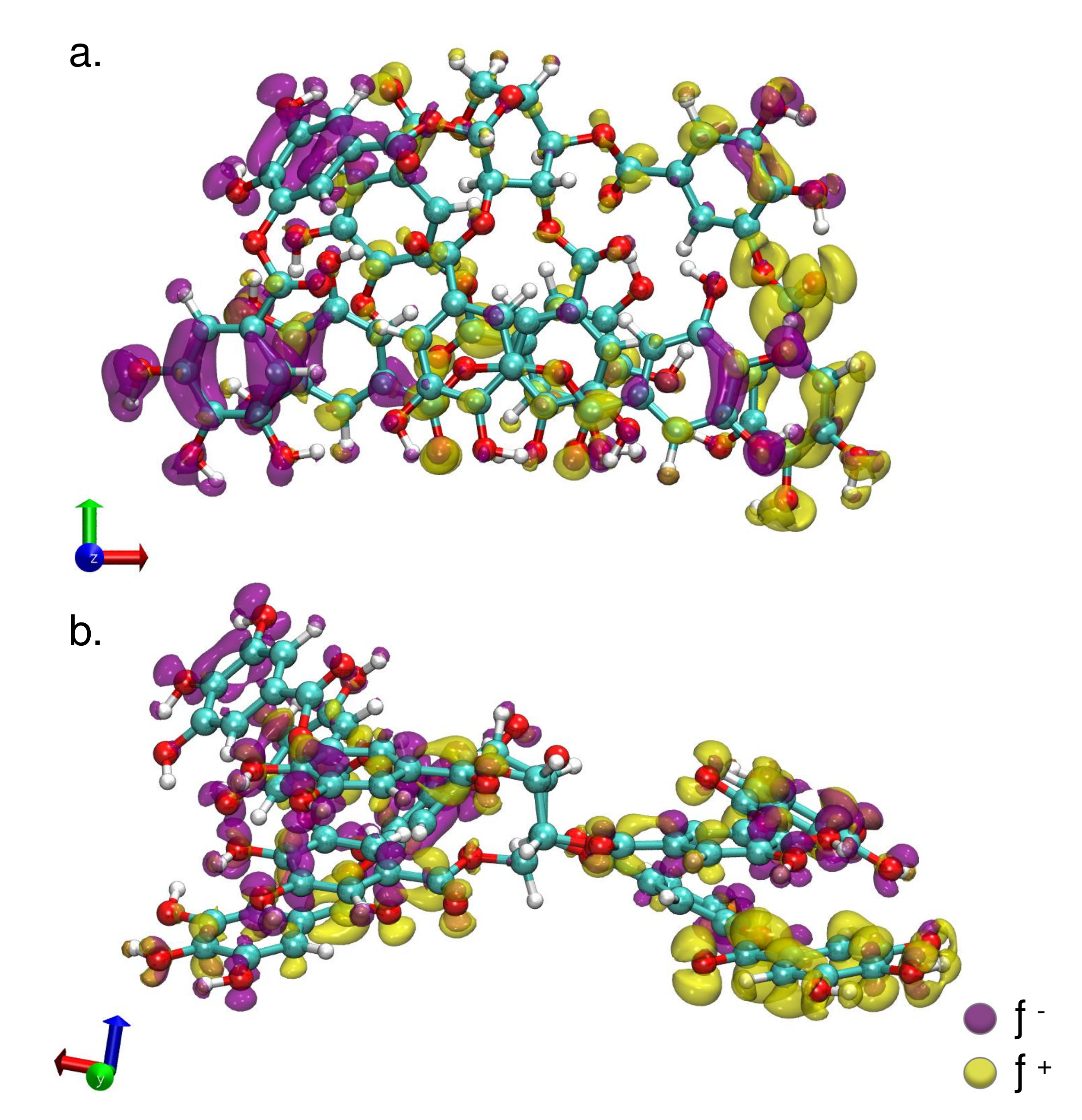}
\caption{Fukui functions $f^+$ and $f^-$, representing the electrophilic and nucleophilic sites of tannic acid structures in (a) vacuum  and (b) water solvent environments. Isosurfaces of $0.13\; e\,\si{\angstrom}^{-3}$.}
\label{fig:fukui}
\end{figure}

\section{Conclusion}
In this work, we demonstrate the external environment effect on tannic acid structure by exploring its conformational free energy surface in gas-phase and considering solvation by water. 
We obtained two sets of different TA geometries (available as Supplementary Material), including the global minima of each. We find that the solvent environment plays a critical role in the structure of TA. While vacuum favors compact structures stabilized by intramolecular interactions between peripheral digallic portions, the molecule in solvent presents more open conformations. The structural changes are also reflected in electronic properties; the frontier molecular orbitals have a significantly increased HOMO-LUMO energy gap in solvent and a different distribution of reactive sites. TA's numerous applications comprehend this molecule's participation in different types of interactions, such as metal coordination, oxidation-reduction (redox) reactions, and the observed TA's structural and electronic changes will have significant effects in these processes. In view of that, our results reinforce the importance of accurate structural and conformational evaluation in determining the interactions of biomolecules such as TA and related polyphenols in the development of new technologies.

\section*{Supplementary Material}
The supplementary material contains principal component analysis maps showing free energies, \new{hierarchical clustering of the ensemble structures,} comparisons between GFN2-xTB, \new{GFN2-xTB including harmonic vibrational free energies  at  298.5 K}, and DFT revPBE-D3(BJ) energies for the lowest energy \new{and representative} conformers, electronic structure of TA conformers, projected density of states of tannic acid structures, binding energy of intramolecular interactions, gallic acid frontier molecular orbitals energies and localization, and condensed Fukui functions of tannic acid structures. We provide the files containing all the structures obtained by the enhanced sampling metadynamics, together with their calculated free energies\new{, and the DFT relaxed structures detailed in Figs. S2 and S3}.

\begin{acknowledgments}
This work is supported by FAPESP (Grants 18/25103-0, 19/04527-0, 17/18139-6, and 17/02317-2). The authors acknowledge the SDumont supercomputer at the Brazilian National Scientific Computing Laboratory (LNCC) for computational resources.
\end{acknowledgments}

\section*{Data Availability}
The data that supports the findings of this study are available within the article and its supplementary material.

%
%



\section*{Supplementary material}

# Supplementary Material

# Conformational analysis of tannic acid: environment effects in electronic and reactivity properties

Romana Petry,*,†,‡ Bruno Focassio,†,‡ Gabriel R. Schleder,†,‡ Diego S. T. Martinez,† and Adalberto Fazzio*,†,‡

†*Brazilian Nanotechnology National Laboratory (LNNano), Brazilian Center for Research in Energy and Materials (CNPEM), Campinas 13083-100, São Paulo, Brazil*

‡*Center for Natural and Human Sciences, Federal University of ABC (UFABC), Santo André, 09210-580, São Paulo, Brazil*

E-mail: romana.petry@lnnano.cnpem.br; adalberto.fazzio@lnnano.cnpem.br

# Energetic evaluation of conformers

## Principal Component Analysis with harmonic free energy colormap

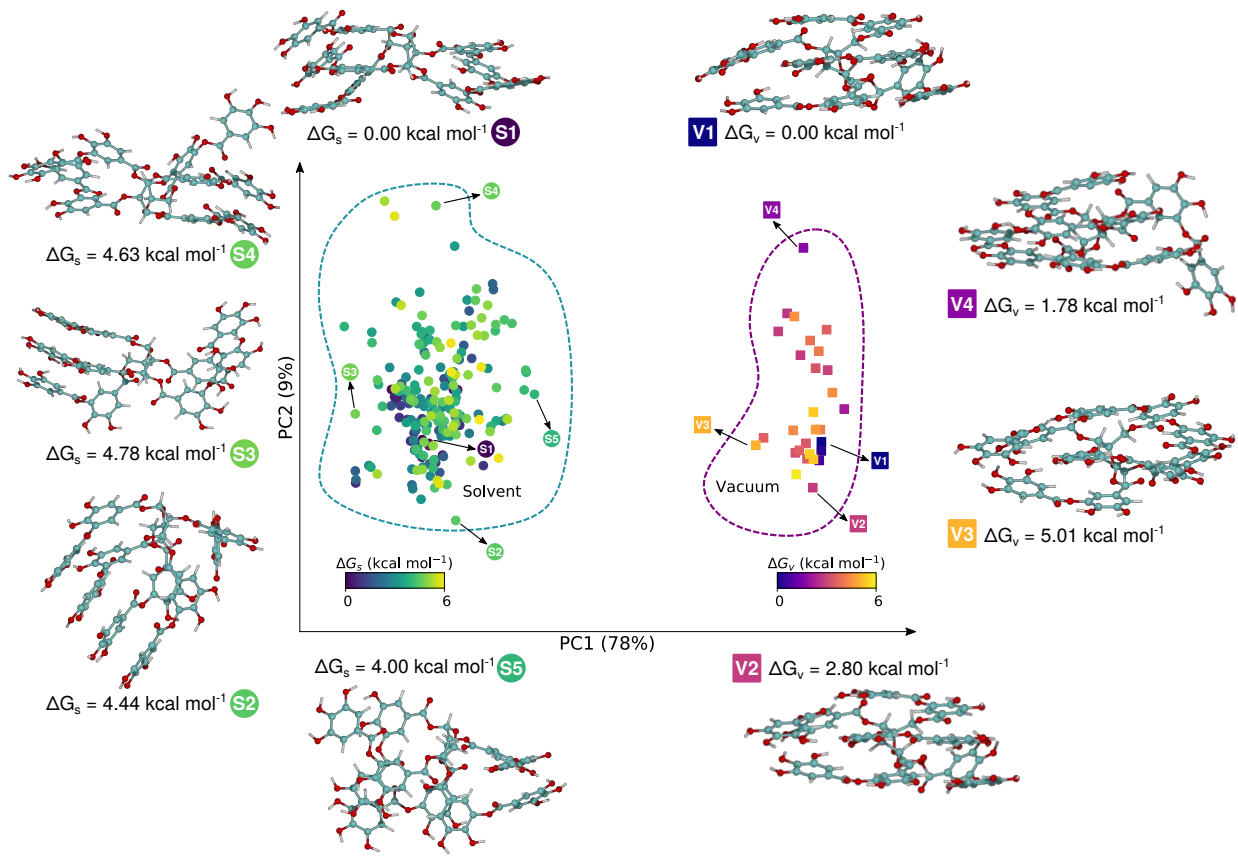

Figure S1: Principal component analysis of conformers configuration space colored by free energy difference. Each square or circle represents a different geometry obtained from vacuum and solvent calculation conditions, respectively. The color scales correspond to the $\Delta G$ of each structure in reference to the global minimum of its respective environment.

# Hierarchical clustering of the ensemble structures

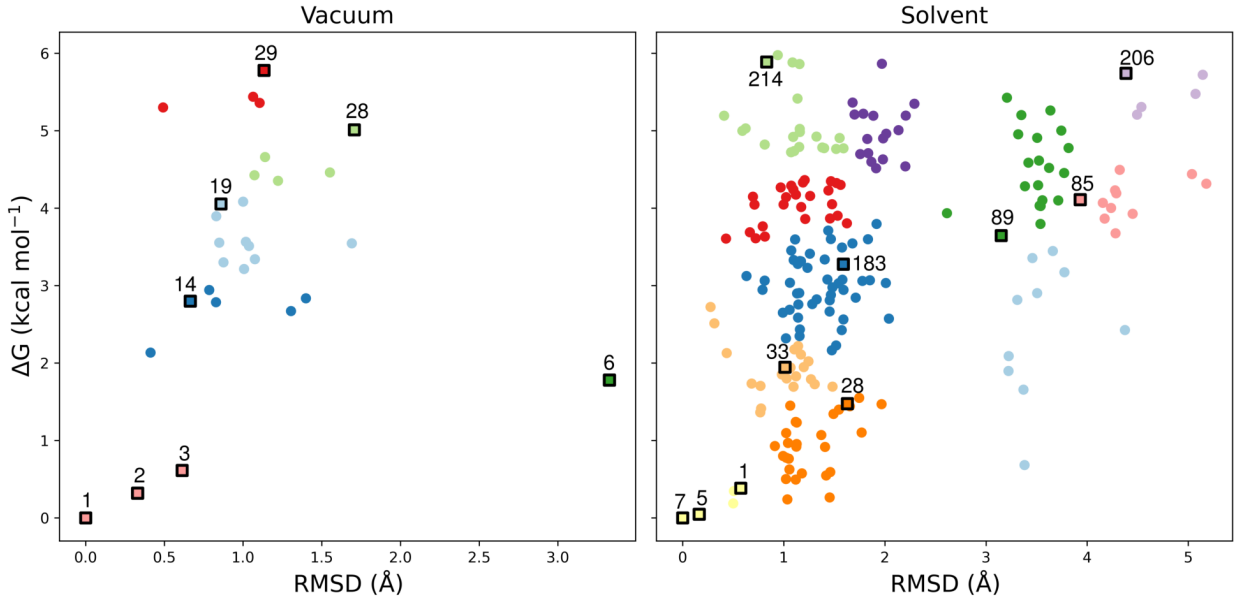

Figure S2: Hierarchical clustering of the ensemble structures obtained in vacuum and water at the xTB level of theory. The structures are grouped based on Ward's agglomerative linkage method, using four features: harmonic vibrational free energy, the structural RMSD, and the two principal components of the PCA of the MBTR descriptor (Figure S1). The different colors represent distinct groups. The structures selected for further electronic calculations are named according to the ascending order of potential energy obtained from each metadynamics calculation.

# GFN2-xTB *vs.* DFT revPBE-D3(BJ)

Table S1: Energy difference of conformers in vacuum and water solvent environments, at the GFN2-xTB and DFT (revPBE-D3(BJ)) levels of theory. The zero reference is given as the energy of the lowest-energy structure obtained at each environment and level of theory.

| Environment | Conformer | $\Delta E$ GFN2-xTB (meV) | $\Delta E$ DFT revPBE-D3(BJ) (meV) |
|---|---|---|---|
| | Conformer 1[a] | **0.0** | +45.8 |
| Vacuum | Conformer 2 | +23.5 | **0.0** |
| | Conformer 3 | +31.0 | +265.9 |
| | Conformer 1[b] | **0.0** | **0.0** |
| Solvent | Conformer 2[c] | +14.0 | +128.3 |
| | Conformer 3 | +12.8 | +157.2 |

[a] potential and free energy minimum
[b] potential energy minimum, structure 7 in Figure S2 (Solvent)
[c] free energy minimum, structure 1 in Figure S2 (Solvent)

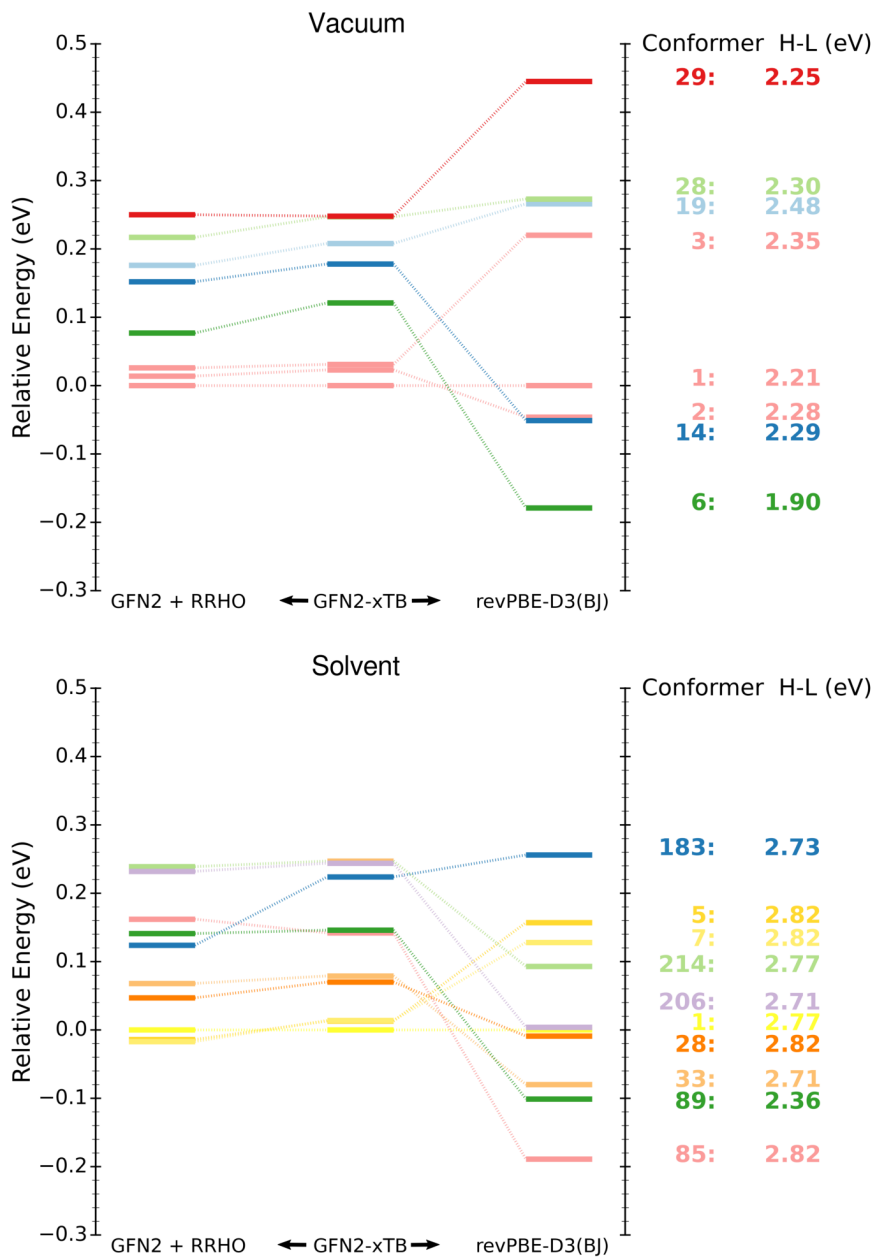

Figure S3: Comparison of conformers (Fig. S2) energy differences obtained between: (middle) GFN2-xTB potential energy, (left) GFN2-xTB including harmonic vibrational free energies at 298.5 K, and (right) DFT total energy employing the revPBE + vdW D3(BJ) functional. Conformer 1 is taken as the zero-energy reference. Free energies were calculated using the harmonic oscillator-rigid rotor (RRHO) approximation at 298.5 K based on the vibrational frequencies calculated at the GFN2-xTB level. The HOMO−LUMO gap (H−L) of each conformer is displayed according to its respective color and DFT energy.

4

# Electronic properties of selected TA structures

## HOMO-LUMO gap and projected density of states of selected conformers from Table S1

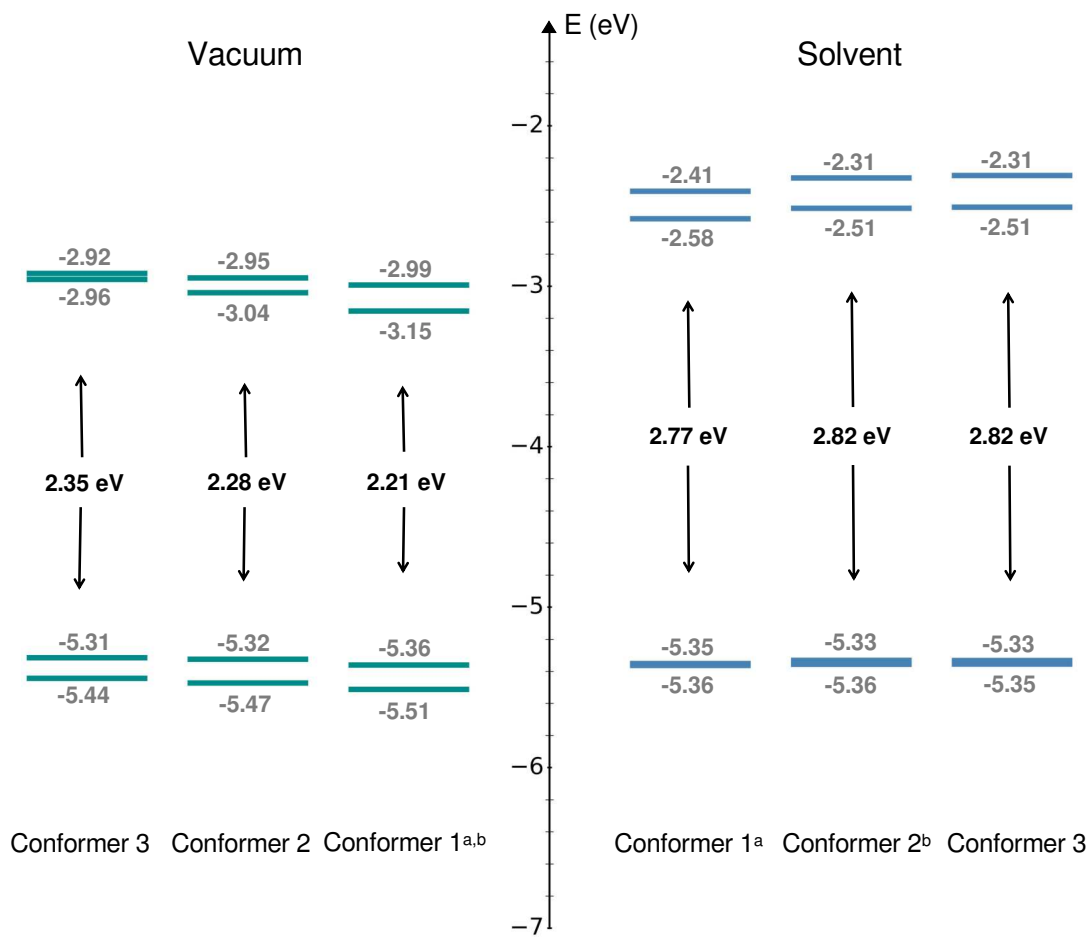

Figure S4: Frontier molecular orbitals of TA conformers obtained in vacuum and implicit solvent: HOMO(−1), HOMO, LUMO, LUMO(+1), and the HOMO-LUMO difference. The representative conformers are the same as those on Table S1, including the potential ($^a$) and free energy ($^b$) minima and their energetically close conformers. The eigenvalues are aligned according to the vacuum level, obtained from the electronic structure calculations of Conformer 1 from the vacuum ensemble.

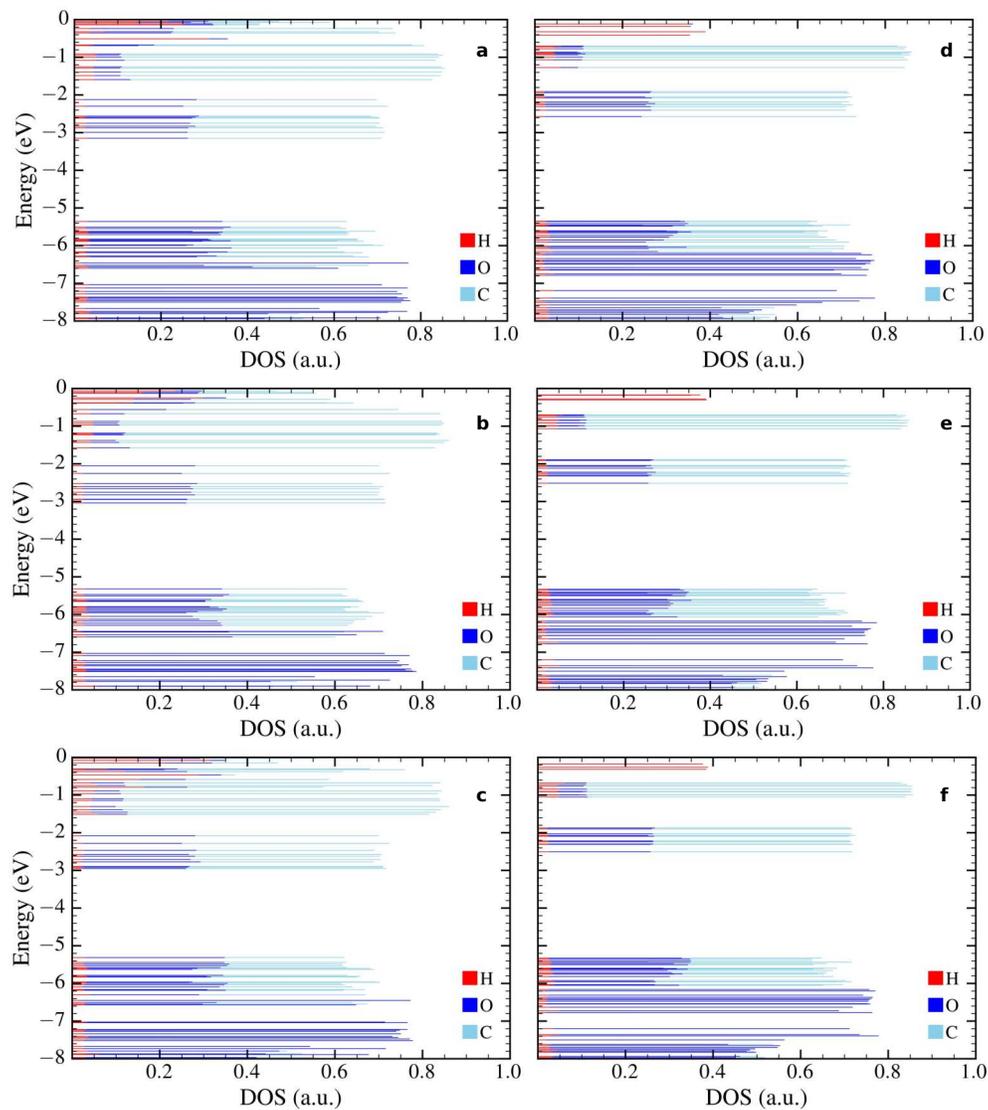

Figure S5: Projected density of states (PDOS) of tannic acid structures obtained in (a,b,c) vacuum and (d,e,f) water solvent. The representative conformers are the same as those on Fig. S4 and Table S1, presented at the same order. The eigenvalues are aligned according to the vacuum level, obtained from the electronic structure calculations of Conformer 1 from the vacuum ensemble.

# Projected density of states and molecular orbitals of selected structures from Figure S2

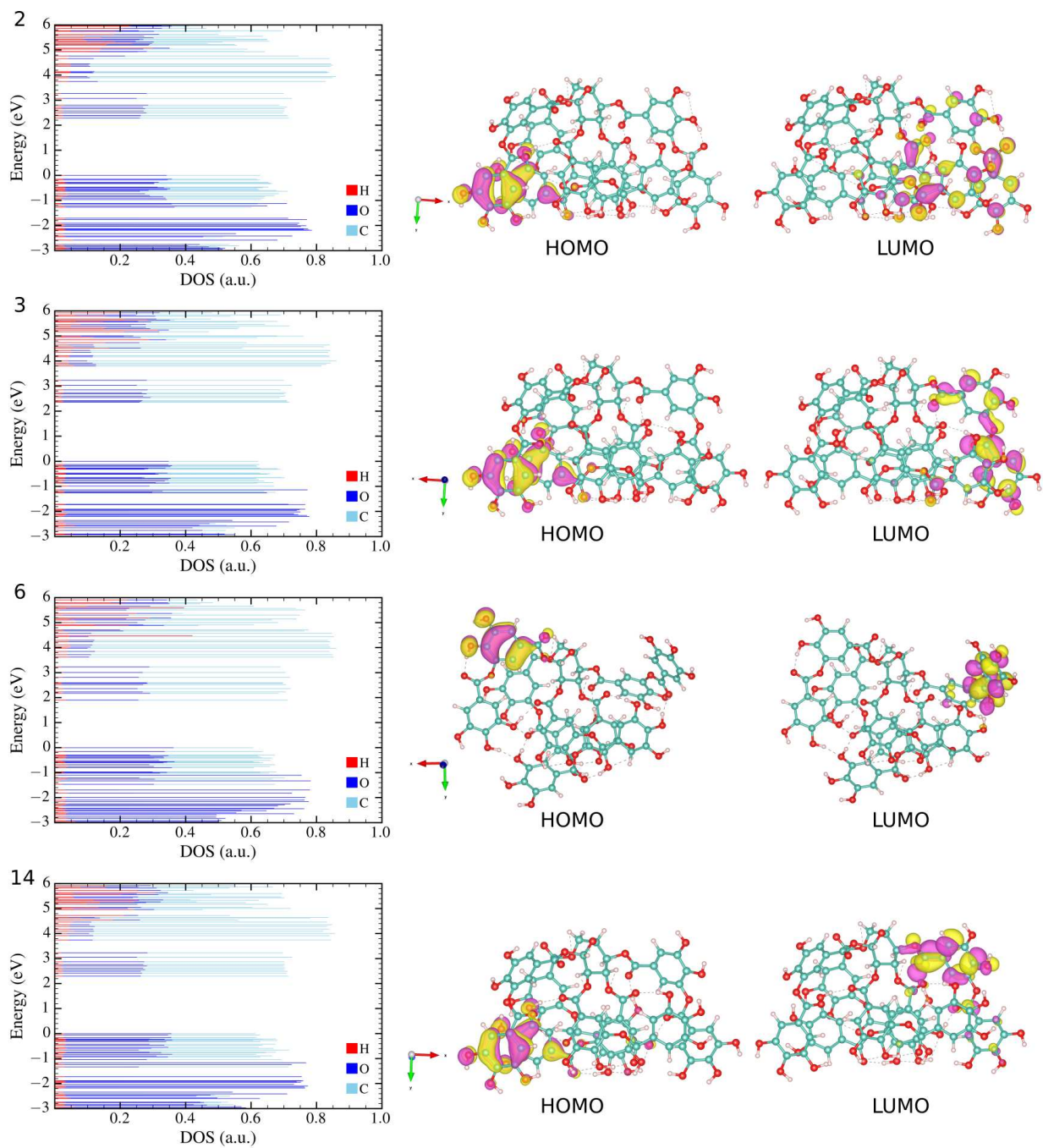

Figure S6: Projected density of states (PDOS) and frontier molecular orbitals (HOMO and LUMO) of tannic acid structures in vacuum. The representative conformers are the same as those on Fig. S2, presented at the same labels. The eigenvalues are aligned according to the HOMO level. Isosurfaces of 0.08 Å$^{-3/2}$. Yellow and purple colors indicate opposite signs of the orbital wave function.

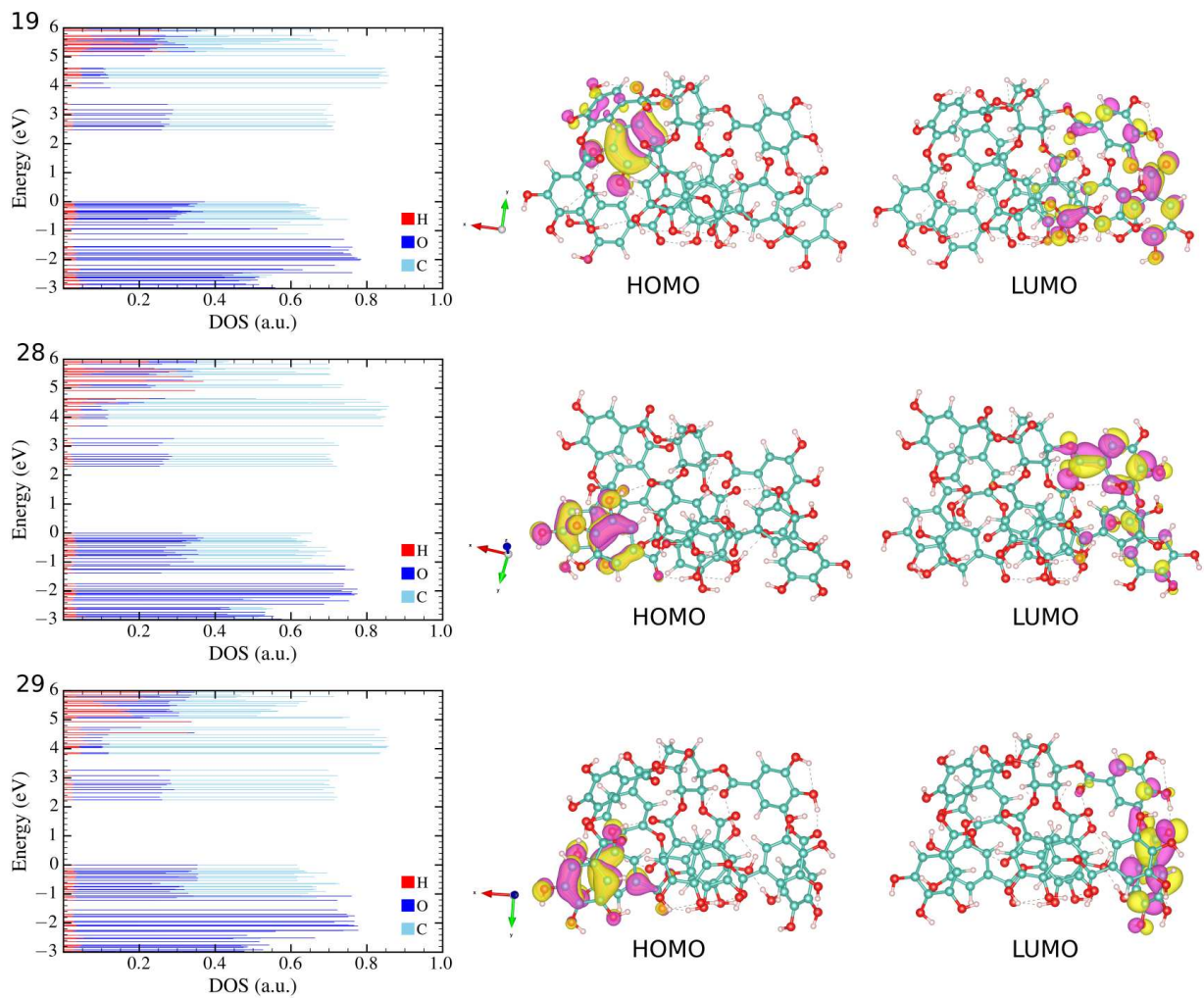

Figure S6 (Cont.): Projected density of states (PDOS) and frontier molecular orbitals (HOMO and LUMO) of tannic acid structures in vacuum. The representative conformers are the same as those on Fig. S2, presented at the same labels. The eigenvalues are aligned according to the HOMO level. Isosurfaces of 0.08 Å$^{-3/2}$. Yellow and purple colors indicate opposite signs of the orbital wave function.

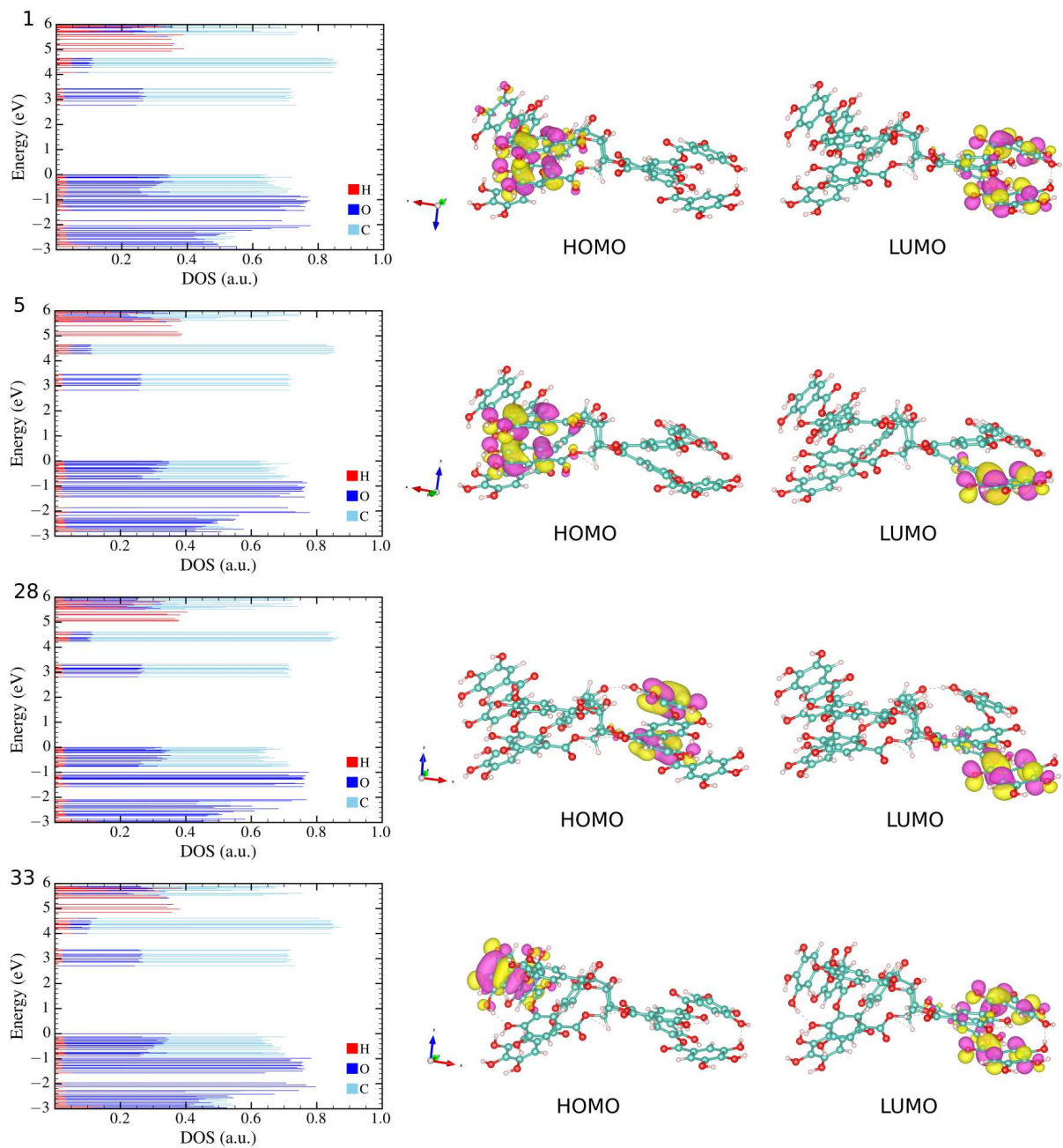

Figure S7: Projected density of states (PDOS) and frontier molecular orbitals (HOMO and LUMO) of tannic acid structures in solvent. The representative conformers are the same as those on Fig. S2, presented at the same labels. The eigenvalues are aligned according to the HOMO level. Isosurfaces of 0.08 Å$^{-3/2}$. Yellow and purple colors indicate opposite signs of the orbital wave function.

9

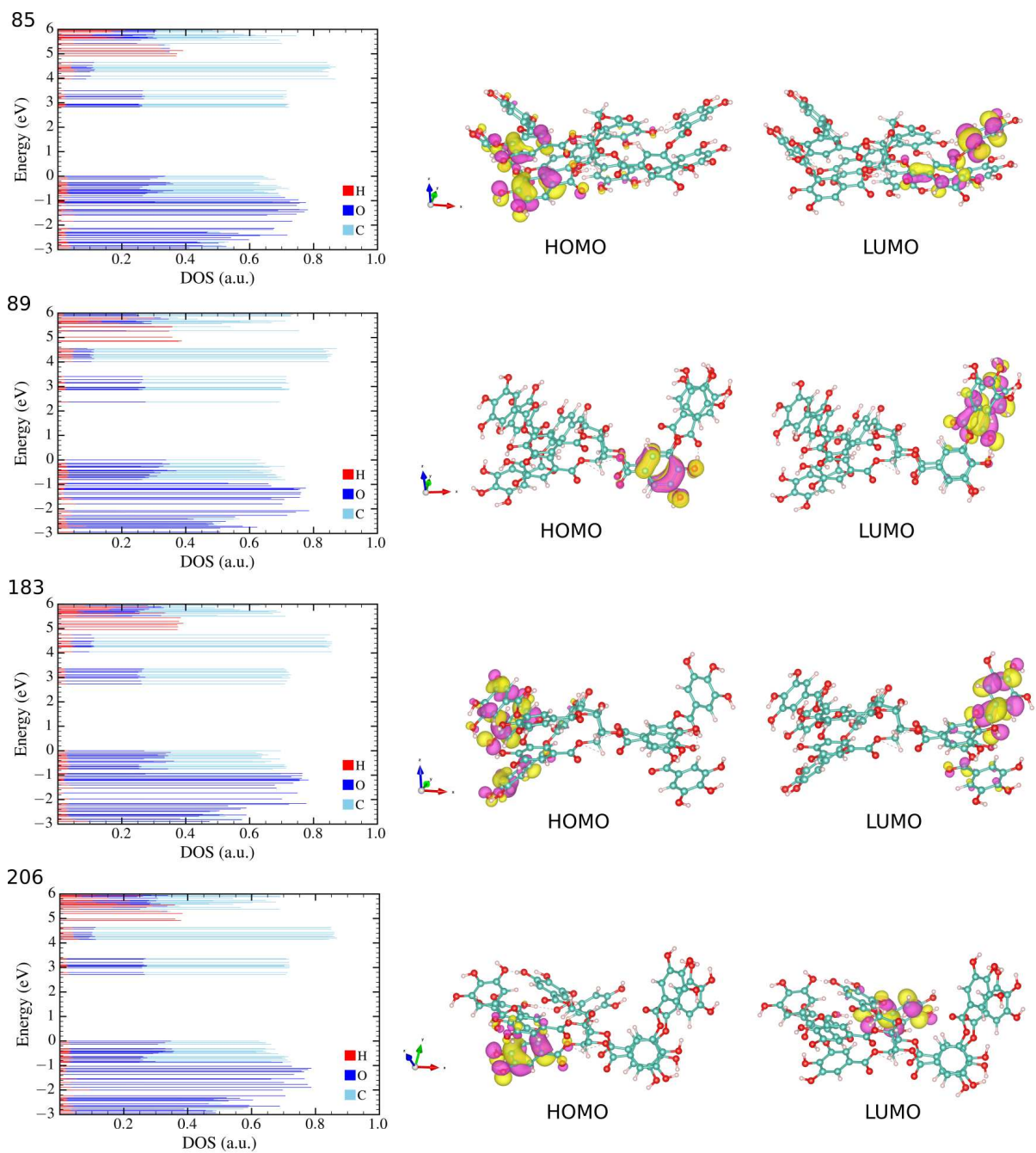

Figure S7 (Cont.): Projected density of states (PDOS) and frontier molecular orbitals (HOMO and LUMO) of tannic acid structures in solvent. The representative conformers are the same as those on Fig. S2, presented at the same labels. The eigenvalues are aligned according to the HOMO level. Isosurfaces of 0.08 Å$^{-3/2}$. Yellow and purple colors indicate opposite signs of the orbital wave function.

214

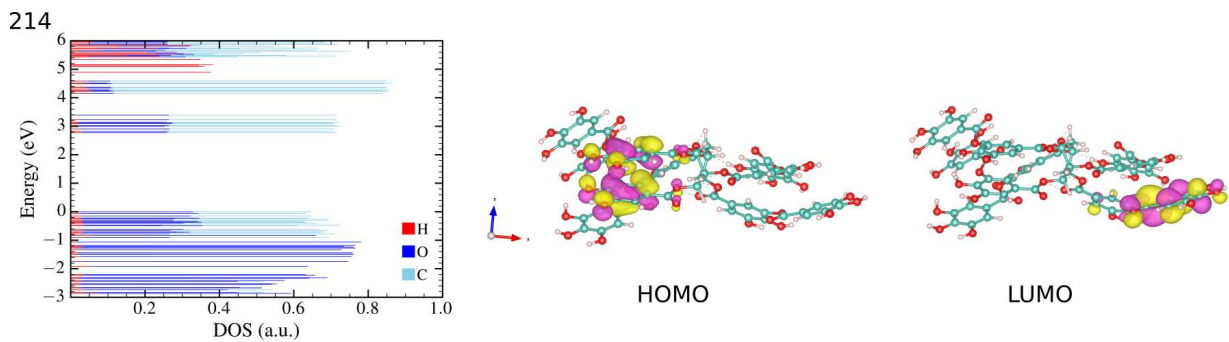

Figure S7 (Cont.): Projected density of states (PDOS) and frontier molecular orbitals (HOMO and LUMO) of tannic acid structures in solvent. The representative conformers are the same as those on Fig. S2, presented at the same labels. The eigenvalues are aligned according to the HOMO level. Isosurfaces of 0.08 Å$^{-3/2}$. Yellow and purple colors indicate opposite signs of the orbital wave function.

# Binding energy of intramolecular interactions

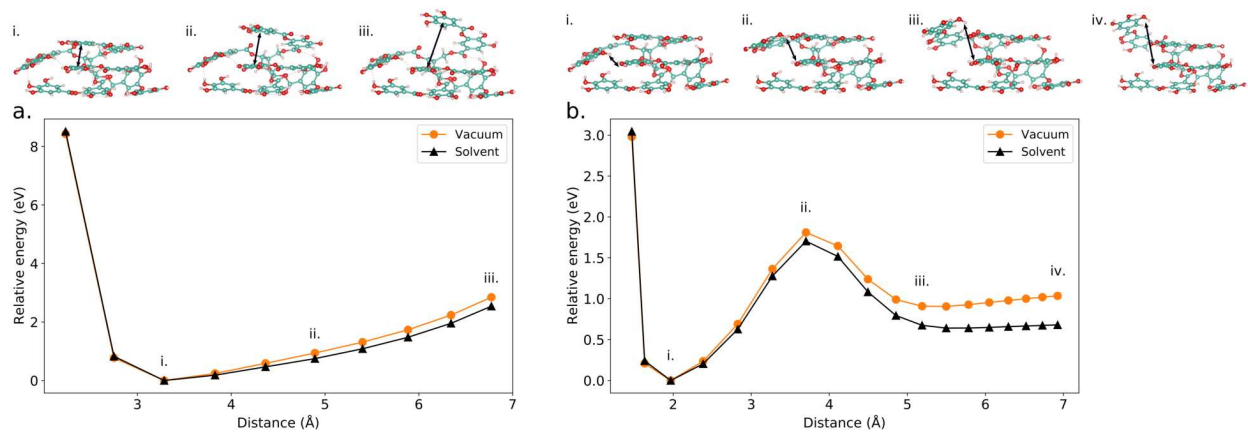

Figure S8: Calculations of tannic acid's energy as a function of distance of a) aromatic functional groups and b) H and O atoms of selected gallic acid portions of the molecule in gas-phase and solvated environment. Structures at different points of the generalized coordinate are represented above each figure. Energies are given in relation to the minimum of each condition. Comparing the intramolecular $\pi$-stacking ($E_b^\pi$) and hydrogen bond ($E_b^H$) interaction energies between the gas-phase and solvated environments, we can estimate that the $E_b^{\pi\,gas-phase} = -2.84$ eV/site decreases to $E_b^{\pi\,solvent} = -2.54$ eV/site, while $E_b^{H\,gas-phase} = -1.03$ eV/site is reduced by almost half in the solvent environment to $E_b^{H\,solvent} = -0.68$ eV/site

# Gallic Acid

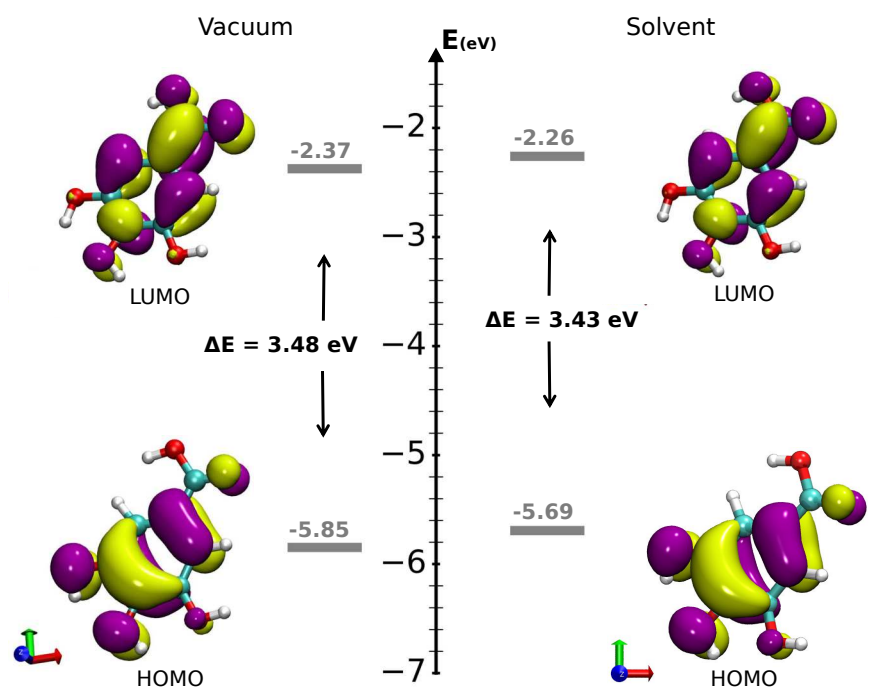

Figure S9: Frontier molecular orbitals and the HOMO-LUMO energy gap of gallic acid structures obtained in vacuum and solvent. The eigenvalues were aligned according to the vacuum level obtained from the optimized vacuum TA calculation. Isosurfaces of 0.08 Å$^{-3/2}$. Yellow and purple colors indicate opposite signs of the orbital wave function.

# Reactivity analysis

Table S2: Condensed Fukui functions ($f^+$ and $f^-$) of tannic acid in vacuum. The atomic charges were defined using Mulliken population analysis. Atom numbers are ordered according to the structure file available as supplementary material.

| \multicolumn{4}{c}{Begin of Table} | | | |
|---|---|---|---|
| Atom | Element | $f^+$ | $f^-$ |
| 1 | C | 0.002 | 0.006 |
| 2 | C | -0.001 | -0.001 |
| 3 | C | 0.004 | 0.003 |
| 4 | C | 0.003 | 0.003 |
| 5 | C | 0.001 | 0.001 |
| 6 | C | 0.000 | 0.003 |
| 7 | C | 0.001 | 0.001 |
| 8 | C | -0.000 | 0.000 |
| 9 | C | 0.000 | 0.000 |
| 10 | C | 0.001 | 0.000 |
| 11 | C | 0.011 | 0.004 |
| 12 | C | 0.010 | 0.005 |
| 13 | C | 0.019 | 0.001 |
| 14 | C | -0.000 | 0.000 |
| 15 | C | 0.004 | 0.003 |
| 16 | C | 0.005 | 0.003 |
| 17 | C | 0.001 | 0.001 |
| 18 | C | 0.001 | 0.004 |
| 19 | C | 0.004 | 0.001 |
| 20 | C | 0.014 | 0.004 |
| 21 | C | 0.003 | 0.005 |

| Continuation of Table S2 ||||
| Atom | Element | $f^+$ | $f^-$ |
| --- | --- | --- | --- |
| 22 | C | 0.002 | 0.001 |
| 23 | C | 0.005 | 0.001 |
| 24 | C | 0.004 | 0.005 |
| 25 | C | 0.003 | 0.003 |
| 26 | C | 0.004 | 0.003 |
| 27 | C | -0.000 | 0.001 |
| 28 | C | -0.000 | 0.003 |
| 29 | C | 0.008 | 0.002 |
| 30 | C | 0.012 | 0.003 |
| 31 | C | 0.003 | 0.002 |
| 32 | C | 0.011 | 0.006 |
| 33 | C | -0.002 | 0.010 |
| 34 | C | 0.009 | 0.001 |
| 35 | C | 0.001 | 0.004 |
| 36 | C | 0.003 | 0.005 |
| 37 | C | 0.007 | 0.003 |
| 38 | C | 0.008 | 0.001 |
| 39 | C | 0.005 | 0.005 |
| 40 | C | 0.002 | 0.007 |
| 41 | C | 0.011 | 0.007 |
| 42 | C | 0.005 | 0.005 |
| 43 | C | 0.001 | 0.014 |
| 44 | C | 0.016 | 0.007 |
| 45 | C | 0.003 | 0.004 |
| 46 | C | 0.014 | 0.002 |

| Continuation of Table S2 | | | |
|---|---|---|---|
| Atom | Element | $f^+$ | $f^-$ |
| 47 | C | 0.003 | 0.005 |
| 48 | C | 0.003 | 0.001 |
| 49 | C | 0.000 | 0.009 |
| 50 | C | 0.002 | 0.009 |
| 51 | C | 0.008 | 0.007 |
| 52 | C | 0.005 | 0.007 |
| 53 | C | 0.001 | 0.006 |
| 54 | C | 0.011 | 0.008 |
| 55 | C | -0.002 | 0.016 |
| 56 | C | 0.006 | 0.008 |
| 57 | C | 0.003 | 0.003 |
| 58 | C | 0.000 | 0.003 |
| 59 | C | -0.000 | 0.011 |
| 60 | C | -0.001 | 0.005 |
| 61 | C | 0.001 | 0.004 |
| 62 | C | 0.002 | 0.009 |
| 63 | C | -0.001 | 0.007 |
| 64 | C | 0.004 | 0.012 |
| 65 | C | 0.009 | 0.007 |
| 66 | C | 0.000 | 0.010 |
| 67 | C | 0.014 | 0.001 |
| 68 | C | 0.006 | 0.001 |
| 69 | C | 0.002 | 0.009 |
| 70 | C | 0.031 | 0.000 |
| 71 | C | 0.004 | -0.001 |

| Continuation of Table S2 |||| 
| Atom | Element | $f^+$ | $f^-$ |
| --- | --- | --- | --- |
| 72 | C | 0.003 | 0.012 |
| 73 | C | 0.018 | 0.002 |
| 74 | C | 0.002 | 0.013 |
| 75 | C | 0.018 | 0.003 |
| 76 | C | 0.008 | 0.003 |
| 77 | H | 0.009 | 0.005 |
| 78 | H | 0.004 | 0.002 |
| 79 | H | -0.000 | 0.000 |
| 80 | H | 0.004 | 0.004 |
| 81 | H | 0.006 | 0.007 |
| 82 | H | 0.006 | 0.003 |
| 83 | H | 0.001 | 0.004 |
| 84 | H | 0.005 | 0.005 |
| 85 | H | 0.010 | 0.006 |
| 86 | H | 0.003 | 0.003 |
| 87 | H | 0.007 | 0.007 |
| 88 | H | 0.003 | 0.003 |
| 89 | H | 0.001 | 0.003 |
| 90 | H | -0.002 | -0.001 |
| 91 | H | 0.005 | 0.007 |
| 92 | H | 0.007 | 0.002 |
| 93 | H | 0.000 | 0.006 |
| 94 | H | 0.002 | 0.001 |
| 95 | H | 0.005 | -0.004 |
| 96 | H | 0.006 | 0.005 |

| Continuation of Table S2 | | | |
|---|---|---|---|
| Atom | Element | $f^+$ | $f^-$ |
| 97 | H | 0.003 | 0.002 |
| 98 | H | 0.003 | 0.001 |
| 99 | H | 0.005 | 0.007 |
| 100 | H | 0.000 | 0.004 |
| 101 | H | 0.010 | 0.012 |
| 102 | H | 0.013 | 0.008 |
| 103 | H | 0.001 | 0.003 |
| 104 | H | -0.002 | 0.000 |
| 105 | H | -0.003 | 0.003 |
| 106 | H | 0.000 | 0.010 |
| 107 | H | 0.004 | 0.005 |
| 108 | H | -0.004 | 0.009 |
| 109 | H | 0.002 | 0.007 |
| 110 | H | 0.002 | 0.002 |
| 111 | H | 0.010 | 0.010 |
| 112 | H | -0.000 | -0.003 |
| 113 | H | 0.012 | 0.002 |
| 114 | H | 0.006 | -0.002 |
| 115 | H | 0.001 | 0.007 |
| 116 | H | 0.003 | 0.002 |
| 117 | H | 0.005 | 0.005 |
| 118 | H | 0.007 | 0.006 |
| 119 | H | 0.002 | 0.009 |
| 120 | H | 0.009 | 0.003 |
| 121 | H | 0.007 | 0.008 |

| Continuation of Table S2 | | | |
|---|---|---|---|
| Atom | Element | $f^+$ | $f^-$ |
| 122 | H | 0.002 | 0.009 |
| 123 | H | 0.001 | 0.006 |
| 124 | H | 0.005 | 0.004 |
| 125 | H | 0.003 | 0.008 |
| 126 | H | 0.005 | 0.013 |
| 127 | H | 0.015 | 0.006 |
| 128 | H | 0.005 | 0.014 |
| 129 | O | 0.009 | 0.008 |
| 130 | O | 0.006 | 0.006 |
| 131 | O | 0.004 | 0.003 |
| 132 | O | 0.000 | 0.000 |
| 133 | O | 0.016 | 0.010 |
| 134 | O | 0.002 | 0.002 |
| 135 | O | 0.009 | 0.011 |
| 136 | O | 0.023 | 0.014 |
| 137 | O | 0.006 | 0.000 |
| 138 | O | 0.006 | 0.003 |
| 139 | O | 0.000 | -0.001 |
| 140 | O | 0.022 | 0.003 |
| 141 | O | 0.007 | 0.004 |
| 142 | O | 0.006 | 0.002 |
| 143 | O | 0.006 | 0.001 |
| 144 | O | -0.000 | 0.004 |
| 145 | O | 0.016 | 0.010 |
| 146 | O | 0.005 | -0.004 |

| Continuation of Table S2 ||||
| Atom | Element | $f^+$ | $f^-$ |
| --- | --- | --- | --- |
| 147 | O | 0.011 | 0.014 |
| 148 | O | 0.004 | 0.014 |
| 149 | O | 0.005 | 0.006 |
| 150 | O | 0.004 | 0.004 |
| 151 | O | -0.000 | 0.007 |
| 152 | O | 0.020 | 0.007 |
| 153 | O | 0.009 | 0.003 |
| 154 | O | 0.008 | 0.008 |
| 155 | O | -0.003 | 0.009 |
| 156 | O | 0.006 | 0.005 |
| 157 | O | 0.005 | 0.014 |
| 158 | O | 0.000 | 0.009 |
| 159 | O | 0.011 | 0.006 |
| 160 | O | 0.011 | 0.007 |
| 161 | O | 0.012 | 0.023 |
| 162 | O | 0.015 | 0.020 |
| 163 | O | 0.005 | 0.019 |
| 164 | O | 0.013 | 0.001 |
| 165 | O | 0.004 | 0.005 |
| 166 | O | 0.015 | 0.019 |
| 167 | O | 0.010 | 0.022 |
| 168 | O | 0.001 | 0.014 |
| 169 | O | 0.009 | 0.021 |
| 170 | O | 0.015 | 0.015 |
| 171 | O | 0.008 | 0.017 |

| Continuation of Table S2 |||| 
| Atom | Element | $f^+$ | $f^-$ |
| --- | --- | --- | --- |
| 172 | O | 0.022 | 0.009 |
| 173 | O | 0.035 | 0.003 |
| 174 | O | 0.007 | 0.031 |

Table S3: Condensed Fukui functions ($f^+$ and $f^-$) of tannic acid in solvent. The atomic charges were defined using Mulliken population analysis. Atom numbers are ordered according to the structure file available as supplementary material.

| Begin of Table | | | |
|---|---|---|---|
| Atom | Element | $f^+$ | $f^-$ |
| 1 | C | -0.001 | -0.000 |
| 2 | C | 0.009 | 0.002 |
| 3 | C | 0.008 | 0.001 |
| 4 | C | -0.002 | -0.001 |
| 5 | C | -0.002 | 0.000 |
| 6 | C | 0.008 | 0.004 |
| 7 | C | 0.019 | 0.005 |
| 8 | C | 0.000 | 0.000 |
| 9 | C | -0.002 | -0.000 |
| 10 | C | -0.000 | -0.000 |
| 11 | C | -0.001 | -0.001 |
| 12 | C | 0.002 | 0.009 |
| 13 | C | 0.003 | 0.008 |
| 14 | C | 0.008 | 0.008 |
| 15 | C | 0.003 | 0.009 |
| 16 | C | 0.009 | 0.008 |
| 17 | C | 0.002 | 0.003 |
| 18 | C | -0.001 | 0.003 |
| 19 | C | 0.003 | 0.009 |
| 20 | C | 0.003 | 0.004 |
| 21 | C | 0.001 | 0.004 |
| 22 | C | 0.011 | 0.008 |

| Continuation of Table S3 | | | |
|---|---|---|---|
| Atom | Element | $f^+$ | $f^-$ |
| 23 | C | 0.000 | 0.005 |
| 24 | C | 0.005 | 0.008 |
| 25 | C | 0.003 | 0.008 |
| 26 | C | 0.002 | 0.006 |
| 27 | C | 0.006 | 0.003 |
| 28 | C | 0.012 | 0.001 |
| 29 | C | 0.005 | 0.005 |
| 30 | C | 0.002 | 0.002 |
| 31 | C | 0.001 | 0.003 |
| 32 | C | 0.002 | 0.006 |
| 33 | C | 0.010 | -0.000 |
| 34 | C | 0.007 | 0.005 |
| 35 | C | 0.001 | 0.004 |
| 36 | C | 0.007 | 0.003 |
| 37 | C | 0.007 | 0.003 |
| 38 | C | 0.007 | 0.007 |
| 39 | C | 0.003 | 0.002 |
| 40 | C | 0.004 | 0.010 |
| 41 | C | 0.003 | 0.005 |
| 42 | C | -0.001 | 0.001 |
| 43 | C | 0.001 | 0.004 |
| 44 | C | 0.001 | 0.004 |
| 45 | C | 0.003 | 0.001 |
| 46 | C | 0.009 | 0.006 |
| 47 | C | 0.003 | 0.005 |

| Continuation of Table S3 | | | |
| --- | --- | --- | --- |
| Atom | Element | $f^+$ | $f^-$ |
| 48 | C | -0.002 | 0.004 |
| 49 | C | 0.015 | 0.002 |
| 50 | C | 0.011 | 0.008 |
| 51 | C | 0.002 | 0.007 |
| 52 | C | 0.007 | 0.006 |
| 53 | C | 0.005 | 0.001 |
| 54 | C | -0.002 | 0.004 |
| 55 | C | 0.005 | 0.004 |
| 56 | C | 0.001 | 0.005 |
| 57 | C | 0.008 | 0.006 |
| 58 | C | -0.001 | 0.006 |
| 59 | C | 0.004 | 0.006 |
| 60 | C | 0.003 | 0.004 |
| 61 | C | 0.026 | 0.001 |
| 62 | C | 0.010 | 0.006 |
| 63 | C | -0.000 | 0.006 |
| 64 | C | 0.004 | 0.003 |
| 65 | C | 0.000 | 0.006 |
| 66 | C | 0.002 | 0.007 |
| 67 | C | 0.011 | 0.009 |
| 68 | C | 0.005 | 0.009 |
| 69 | C | 0.008 | 0.003 |
| 70 | C | 0.001 | 0.003 |
| 71 | C | 0.001 | 0.007 |
| 72 | C | 0.008 | 0.002 |

| Continuation of Table S3 |||| 
|---|---|---|---|
| Atom | Element | $f^+$ | $f^-$ |

| Atom | Element | $f^+$ | $f^-$ |
|---|---|---|---|
| 73 | C | 0.005 | 0.005 |
| 74 | C | 0.016 | 0.003 |
| 75 | C | 0.017 | 0.002 |
| 76 | C | 0.008 | 0.002 |
| 77 | H | 0.002 | -0.000 |
| 78 | H | 0.007 | 0.007 |
| 79 | H | 0.002 | 0.007 |
| 80 | H | 0.007 | 0.007 |
| 81 | H | 0.009 | 0.004 |
| 82 | H | -0.003 | -0.002 |
| 83 | H | 0.008 | 0.005 |
| 84 | H | 0.011 | 0.005 |
| 85 | H | 0.007 | 0.006 |
| 86 | H | 0.002 | 0.002 |
| 87 | H | 0.007 | 0.008 |
| 88 | H | 0.003 | 0.003 |
| 89 | H | 0.007 | 0.008 |
| 90 | H | 0.006 | 0.005 |
| 91 | H | 0.001 | 0.003 |
| 92 | H | -0.003 | 0.002 |
| 93 | H | 0.009 | 0.008 |
| 94 | H | 0.005 | 0.010 |
| 95 | H | 0.002 | 0.003 |
| 96 | H | 0.008 | 0.007 |
| 97 | H | 0.006 | 0.010 |

| Continuation of Table S3 | | | |
|---|---|---|---|
| Atom | Element | $f^+$ | $f^-$ |
| 98 | H | 0.005 | 0.003 |
| 99 | H | 0.005 | 0.005 |
| 100 | H | 0.005 | 0.009 |
| 101 | H | 0.001 | 0.002 |
| 102 | H | -0.001 | 0.002 |
| 103 | H | 0.007 | 0.011 |
| 104 | H | 0.005 | 0.007 |
| 105 | H | 0.006 | 0.006 |
| 106 | H | 0.007 | 0.005 |
| 107 | H | 0.004 | 0.008 |
| 108 | H | 0.006 | 0.008 |
| 109 | H | 0.010 | 0.007 |
| 110 | H | 0.004 | 0.006 |
| 111 | H | 0.005 | 0.003 |
| 112 | H | 0.002 | 0.003 |
| 113 | H | 0.006 | 0.008 |
| 114 | H | 0.007 | 0.010 |
| 115 | H | 0.006 | 0.009 |
| 116 | H | -0.001 | 0.002 |
| 117 | H | 0.007 | 0.008 |
| 118 | H | 0.014 | 0.005 |
| 119 | H | 0.002 | 0.006 |
| 120 | H | 0.006 | 0.008 |
| 121 | H | 0.002 | 0.007 |
| 122 | H | 0.007 | 0.008 |

| Continuation of Table S3 ||||
| Atom | Element | $f^+$ | $f^-$ |
| --- | --- | --- | --- |
| 123 | H | 0.003 | 0.003 |
| 124 | H | 0.004 | 0.003 |
| 125 | H | 0.001 | 0.003 |
| 126 | H | 0.006 | 0.007 |
| 127 | H | 0.009 | 0.000 |
| 128 | H | 0.005 | 0.007 |
| 129 | O | 0.019 | 0.014 |
| 130 | O | 0.004 | 0.001 |
| 131 | O | 0.018 | 0.012 |
| 132 | O | 0.003 | 0.000 |
| 133 | O | 0.011 | 0.015 |
| 134 | O | -0.003 | -0.001 |
| 135 | O | 0.011 | 0.008 |
| 136 | O | 0.017 | 0.010 |
| 137 | O | 0.005 | 0.005 |
| 138 | O | 0.007 | 0.005 |
| 139 | O | -0.001 | 0.001 |
| 140 | O | 0.018 | 0.007 |
| 141 | O | 0.002 | 0.001 |
| 142 | O | 0.005 | 0.013 |
| 143 | O | 0.018 | 0.013 |
| 144 | O | 0.015 | 0.006 |
| 145 | O | 0.005 | 0.012 |
| 146 | O | 0.005 | 0.003 |
| 147 | O | 0.005 | 0.016 |

| Continuation of Table S3 | | | |
|---|---|---|---|
| Atom | Element | $f^+$ | $f^-$ |
| 148 | O | 0.007 | 0.012 |
| 149 | O | 0.004 | 0.003 |
| 150 | O | 0.024 | 0.009 |
| 151 | O | 0.005 | 0.011 |
| 152 | O | -0.001 | -0.001 |
| 153 | O | 0.005 | 0.006 |
| 154 | O | 0.008 | 0.015 |
| 155 | O | 0.007 | 0.006 |
| 156 | O | 0.005 | 0.009 |
| 157 | O | 0.007 | 0.011 |
| 158 | O | 0.039 | 0.006 |
| 159 | O | 0.005 | 0.012 |
| 160 | O | 0.003 | 0.004 |
| 161 | O | 0.007 | 0.001 |
| 162 | O | 0.009 | 0.014 |
| 163 | O | 0.007 | 0.012 |
| 164 | O | 0.004 | 0.016 |
| 165 | O | 0.004 | 0.009 |
| 166 | O | 0.011 | 0.014 |
| 167 | O | 0.010 | 0.009 |
| 168 | O | 0.004 | 0.010 |
| 169 | O | 0.008 | 0.007 |
| 170 | O | 0.005 | 0.016 |
| 171 | O | 0.003 | 0.006 |
| 172 | O | 0.004 | 0.012 |

| Continuation of Table S3 ||||
|---|---|---|---|
| Atom | Element | $f^+$ | $f^-$ |
| 173 | O | 0.006 | 0.011 |
| 174 | O | 0.012 | 0.005 |



\end{document}